\documentclass[a4paper,12pt]{article}
\usepackage{amssymb}
\usepackage{amsmath}
\usepackage{mathtools}
\usepackage{slashed}
\usepackage{color}
\usepackage{indentfirst}
\usepackage{graphicx}
\usepackage{csquotes} 
\usepackage{caption}
\usepackage{comment}
\usepackage{cite}

\renewcommand{\thefootnote}{\fnsymbol{footnote}}

\begin{document}

\title{
\begin{flushright}
\begin{minipage}{0.2\linewidth}
\normalsize
EPHOU-20-011\\
KEK-TH-2265 \\
WU-HEP-20-10\\*[50pt]
\end{minipage}
\end{flushright}
{\Large \bf 
Spontaneous CP violation and symplectic modular symmetry 
in Calabi-Yau compactifications
\\*[20pt]}}

\author{Keiya Ishiguro$^{a}$\footnote{
E-mail address: keyspire@ruri.waseda.jp
},\,
Tatsuo Kobayashi$^{b}$\footnote{
E-mail address: kobayashi@particle.sci.hokudai.ac.jp
}
\ and\
Hajime~Otsuka$^{c}$\footnote{
E-mail address: hotsuka@post.kek.jp
}\\*[20pt]
$^a${\it \normalsize 
Department of Physics, Waseda University, Tokyo 169-8555, Japan} \\
$^b${\it \normalsize 
Department of Physics, Hokkaido University, Sapporo 060-0810, Japan} \\
$^c${\it \normalsize 
KEK Theory Center, Institute of Particle and Nuclear Studies, KEK,}\\
{\it \normalsize 1-1 Oho, Tsukuba, Ibaraki 305-0801, Japan}}
\maketitle

\date{
\centerline{\small \bf Abstract}
\begin{minipage}{0.9\linewidth}
\medskip 
\medskip 
\small
We explore the geometrical origin of CP and the spontaneous CP violation in Calabi-Yau compactifications. We find that the CP symmetry is identified with an outer automorphism of the symplectic modular group in the large complex structure regime of Calabi-Yau threefolds, thereby enlarging the symplectic modular group to their semidirect product group. 
The spontaneous CP violation is realized by the introduction of fluxes, whose effective  action is invariant under CP as well as the discrete $\mathbb{Z}_2$ symmetry or $\mathbb{Z}_4$ R-symmetry. 
We explicitly demonstrate the spontaneous CP violation on a specific Calabi-Yau threefold.
\end{minipage}
}

\renewcommand{\thefootnote}{\arabic{footnote}}
\setcounter{footnote}{0}
\thispagestyle{empty}
\clearpage
\addtocounter{page}{-1}

\tableofcontents

\section{Introduction}
\label{sec:1}

The origin of CP violation is one of important issues to study in particle physics.
Indeed, it has been studied extensively in various scenarios.
CP violation may originate from an underlying theory such as superstring theory.

As discussed in Refs.~\cite{Strominger:1985it,Dine:1992ya,Choi:1992xp},  
one can embed the four-dimensional (4D) CP symmetry into 
proper Lorentz symmetry in higher dimensional theory, e.g., 
ten-dimensional (10D) proper Lorentz symmetry in superstring theory 
on six-dimensional (6D) Calabi-Yau (CY) manifolds.
That is, as the 10D proper Lorentz transformation, 
one performs simultaneously the 4D spacetime orientation and orientation reversing transformation of CY threefolds.
The latter 6D transformation can be achieved by $z_i \rightarrow{-z^\ast_i}$ with $z_i$, $i=1,2,3$, being the complex coordinates, corresponding to the negative determinant in  the transformation of the 6D CY manifolds.
(See also for CP symmetry on 6D orbifold compactifications \cite{Kobayashi:1994ks}.) 
The 6D transformation reverses the sign of the volume form as well as the K\"ahler form of CY threefolds. 
To clarify the nature of CP symmetry, let us consider Majorana-Weyl spinor representation of 10D Lorentz group $SO(1,9)$ which decomposes as ${\bf 16}=({\bf 2},{\bf 4})\oplus ({\bf 2'}, {\bf\bar{4}})$ under 
the 4D and the 6D Lorentz groups
$SO(1,3)\times SO(6)$. 
Here, ${\bf 2}$ and ${\bf 2'}$ denote the left- and right-handed spinor representations of $SL(2,\mathbb{C})$, 
and ${\bf 4}$ and ${\bf \bar{4}}$ represent the positive- and negative-chirality spinor representations of $SO(6)\simeq SU(4)$, 
respectively. 
Then, simultaneous transformations of 4D parity and 6D  orientation reversing exchange 
$({\bf 2}, {\bf 4})$ with $({\bf 2'}, {\bf \bar{4}})$. 
For example, in the context of standard embedding 
in the heterotic string theory, fundamental and 
anti-fundamental representations of $E_6$, ${\bf 27}$ and ${\bf \overline{27}}$, correspond to ${\bf 4}$ and ${\bf \bar{4}}$, 
respectively. 
Hence, they are exchanged under such simultaneous transformations. 
Such a CP transformation is identified with an outer 
automorphism transformation of gauge symmetries~\cite{Grimus:1995zi} and 
each of 4D and 6D Lorentz symmetries~\cite{Buchbinder:2000cq}.

In Type II D-brane models, 
massless fermions localize on D-branes wrapping certain cycles of the CY threefold. 
When we consider the orientation reversing transformation on the internal cycles wrapped by D-branes, 
one can also embed the 4D CP into the higher-dimensional Lorentz transformation on the worldvolume of D-branes on which CP exchanges 
left-handed massless matters with their complex conjugates.

The torus compactification as well as toroidal orbifold compactifications 
is one of simplest compactifications.
The 2D torus has two independent one-cycles.
We have the degree of freedom to change bases of these cycles.
That is the $SL(2, \mathbb{Z})$ transformation, which is called as 
the modular symmetry.
The modular group is the geometrical symmetry,  and also 
transforms zero-modes corresponding to quarks and leptons, 
that is, the flavor symmetry.
(See for the modular symmetry in magnetized D-brane models \cite{Kobayashi:2018rad,Kikuchi:2020frp,Kikuchi:2020nxn} and the classification of 
its subgroup \cite{Kobayashi:2020hoc}.)\footnote{
Recently, the modular flavor symmetries are studied extensively from
the phenomenological viewpoint \cite{Feruglio:2017spp}.}
When there are 4D models having non-trivial flavor symmetries, 
the 4D CP symmetry can be enlarged to a generalized CP symmetry, 
which includes outer automorphisms of flavor symmetries \cite{Feruglio:2012cw,Holthausen:2012dk,Chen:2014tpa}.
Such a generalized CP symmetry is also extended to 
the modular symmetry on the torus compactification, where 
CP symmetry is identified with the outer automorphism of $SL(2, \mathbb{Z})$ \cite{Nilles:2018wex,Novichkov:2019sqv}.

The size and shape of a compact space is described by moduli parameters, 
which are vacuum expectation values of moduli fields.
Their vacuum expectation values can be determined at the potential minima of moduli fields.
The complex structure moduli also transform under the CP symmetry, 
$z_i \rightarrow{-z^\ast_i}$.
Thus, the CP symmetry can be spontaneously broken 
through the moduli stabilization.
For example, the spontaneous CP-violation was studied by assuming 
non-perturbative moduli potentials \cite{Dent:2001cc,Khalil:2001dr,Giedt:2002ns,Kobayashi:2019uyt}.
The three-form flux background is of controllable ways to stabilize 
the complex structure moduli as well as  the axio-dilaton \cite{Gukov:1999ya}.
Indeed, in Ref.~\cite{Kobayashi:2020uaj}, spontaneous CP-violation was studied 
within the framework of Type IIB superstring theory on
toroidal orbifold compactifications with three-form flux background.
The CP-invariant superpotential, which is induced by three-form fluxes, 
is either even or odd polynomial functions of the complex structure moduli 
and the axio-dilaton.
The potential minima were examined, but 
the spontaneous CP-violation can not be realized in 
the torus compactifications.

Our purpose in this paper is to reveal the geometrical origin of CP and search for the spontaneous CP violation 
in CY compactifications, with an emphasis on 
the effective action of the complex structure moduli. 
As discussed in Ref.~\cite{Strominger:1985it}, the orientation reversing isometry of CY threefolds 
induces anti-holomorphic transformations for the complex structure, 
which restrict the form of low-energy effective action. 
We find that CP symmetry is identified with the outer automorphism of the symplectic modular group of generic CY 
threefolds in the large complex structure regime. 
That is a natural extension of the discussion in toroidal backgrounds, where the 
CP symmetry is 
also identified with the outer automorphism of $SL(2, \mathbb{Z})$ modular group on the 
2D torus. 

Furthermore, we discuss the spontaneous CP violation in the CY flux compactifications. 
Flux compactifications in the string theory have a potential to break CP spontaneously. 
In a similar discussion with the toroidal background~\cite{Kobayashi:2020uaj}, the flux-induced superpotential is restricted to be odd or even polynomials with respect to the moduli fields, having the discrete $\mathbb{Z}_2$ symmetry or $\mathbb{Z}_4$ R-symmetry 
from the field theoretical point of view. 
It turns out that the CP-conserving vacua exist in the large complex structure regime of generic CY threefolds. 
The spontaneous CP violation occurs in some class of CY threefolds whose prepotentials have a structure 
different from the toroidal one. 

This paper is organized as follows. 
In Sec.~\ref{sec:2}, we show a relationship between the CP symmetry and the symplectic modular symmetry. 
Sec.~\ref{sec:3} is devoted to the construction of the flux-induced effective action in a CP-invariant way. 
Concrete CY flux compactifications are analyzed in Sec.~\ref{sec:4}, in which the CP-breaking and -conserving 
vacua are demonstrated. 
Finally, we conclude our results in Sec.~\ref{sec:con}. 

\section{CP and symplectic modular symmetries}
\label{sec:2}

In this section, we focus on the complex structure moduli of CY threefolds ${\cal M}$, 
whose effective action is described by the K\"ahler potential in the reduced Planck mass unit $M_{\rm Pl}=1$,
\begin{align}
    K_{\rm cs} &= -\ln \biggl[-i \int_{{\cal M}} \Omega \wedge \bar{\Omega}\biggl],
\label{eq:K}
\end{align}
where $\Omega$ denotes a holomorphic three-form of CY threefolds determining the complex structure 
of ${\cal M}$. 
We examine the CP invariance of the holomorphic three-form $\Omega$ as well as its 
relation to the symplectic modular symmetry on ${\cal M}$.

Recalling that $i\Omega \wedge \bar{\Omega}$ is represented by the volume form of CY threefolds $dV$, namely $i\Omega \wedge \bar{\Omega} = ||\Omega||^2 dV$ with $||\Omega||^2$ being a scalar, 
the orientation reversing transformation changes  the sign of the volume form leading 
$i\Omega \wedge \bar{\Omega} \rightarrow -i\Omega \wedge \bar{\Omega}$. 
It results in the transformation of $\Omega$ under 
the orientation reversing\footnote{Although the following discussion also holds for the anti-holomorphic involution $\Omega \rightarrow \bar{\Omega}$ corresponding to $z_i\rightarrow \bar{z}_i$ in the local patch, 
we adopt the orientation reversing in 
Eq.~(\ref{eq:Omegatrf}) without loss of generality.},
\begin{align}
    \Omega \rightarrow{-\bar{\Omega}}.
    \label{eq:Omegatrf}
\end{align}
Note that in the local coordinates of CY threefolds $\{z_i\}$, the holomorphic three-form is given by 
$\Omega= dz_1\wedge dz_2 \wedge dz_3$. Hence, the orientation reversing transformation 
$z_i \rightarrow -\bar{z}_i$ gives rise to Eq.~(\ref{eq:Omegatrf}).

Since CY threefolds are described by the special geometry\footnote{For more details about the special geometry, 
we refer Refs.~\cite{Strominger:1990pd,Candelas:1990pi}.}, 
the holomorphic three-form is expanded in the symplectic basis. 
When we denote $(A^I, B_I)$ ($I,J=0,1,\cdots, h^{2,1}({\cal M})$) a canonical homology basis for $H_3({\cal M},\mathbb{Z})$, the dual cohomology basis ($\alpha_I,\beta^I$) is defined such that
\begin{align}
    \int_{A^J}\alpha_I=-\int_{B_I}\beta^J=\int_{{\cal M}} \alpha_I \wedge \beta^J =\delta^J_I,\quad
    \int_{{\cal M}} \alpha_I \wedge \alpha_J = \int_{{\cal M}} \beta^I \wedge \beta^J =0.
\end{align}
In terms of the real three-form basis ($\alpha_I,\beta^I$), the holomorphic three-form can be expanded as
\begin{align}
    \Omega &= X^I \alpha_I -{\cal F}_I\beta^I,
\label{eq:Omegaexpand}
\end{align}
with ${\cal F}_I \equiv \partial_I {\cal F}$. Here, ${\cal F}$ is the prepotential as a function of homogeneous coordinates $X^I$ on the moduli space. 
Note that we have rescaling degrees of freedom on $X^I$.

Let us analyze the CP transformation of $\Omega$ in the symplectic basis in more detail. 
The transformation of $\Omega$ in Eq.~(\ref{eq:Omegatrf}) 
is satisfied when 
\begin{align}
    X^0\alpha_0 \rightarrow{-\bar{X}^0\alpha_0},\quad
    X^i\alpha_i \rightarrow{-\bar{X}^i\alpha_i},\quad
    {\cal F}_0\beta^0 \rightarrow{-\bar{{\cal F}}_0\beta^0},\quad
    {\cal F}_i\beta^i \rightarrow{-\bar{{\cal F}}_i\beta^i}.
    \label{eq:CPtrf}
\end{align}
We define the flat coordinates $u^i =X^i/X^0$, $i=1,2,\cdots, h^{2,1}$.
The orientation reversing transformation requires that the complex structure moduli should transform $u^i \rightarrow{\pm\bar{u}^i}$. 
In the following discussion, we adopt $u^i \rightarrow -\bar{u}^i$, restricting ourselves to the ${\rm Im}u^i>0$ plane\footnote{
That is a generalization of the upper half plane of the complex plane, realized in 
the $SL(2, \mathbb{Z})$ moduli space of the torus.} and focus on 
the large complex structure regime. 
Note that the space of harmonic forms splits under the CP transformation into even and odd eigenspaces, for instance, 
\begin{align}
     H^3({\cal M}) = H^{3({\rm CP})}_+({\cal M}) +  H^{3({\rm CP})}_-({\cal M}),
\end{align}
for the third cohomology group. 
The CP-even and -odd bases are elements of
\begin{align}
     H^{3({\rm CP})}_+({\cal M})=\{\alpha_{M}^{({\rm CP})}, \beta^{({\rm CP})\tilde{M}}\},
     \quad
     H^{3({\rm CP})}_-({\cal M})=\{\alpha_{\tilde{M}}^{({\rm CP})}, \beta^{({\rm CP})M}\},
\end{align}
with $M,\tilde{M} = 0,1,...,h^{2,1}$, whose intersections satisfy
\begin{align}
     \int_{{\cal M}} \alpha_{N}^{({\rm CP})} \wedge \beta^{({\rm CP})M} =\delta^M_N,
     \quad
     \int_{{\cal M}} \alpha_{\tilde{N}}^{({\rm CP})} \wedge \beta^{({\rm CP})\tilde{M}} =  \delta^{\tilde{M}}_{\tilde{N}},
     \label{eq:intersection}
\end{align}
otherwise 0. The definition of CP-even/odd bases are analogous to Type IIA orientifolds with O6-planes \cite{Grimm:2004ua} and also consistent with orientifold projection in Type IIB string theory with O3/O7-planes \cite{Grimm:2004uq} as discussed in Sec. \ref{sec:3_1}. These structures can be seen in the local patch parametrized by $z_i = x_i +iy_i$. Indeed, CP-even and odd bases under the orientation reversing $\{x_i\rightarrow -x_i,  y_i\rightarrow y_i\}$ are locally given by 
$\{dx_i \wedge dx_j \wedge dy_k, dy_i \wedge dy_j \wedge dy_k\}$ and $\{dx_i \wedge dx_j \wedge dx_k, dx_i \wedge dy_j \wedge dy_k\}$, $i\neq j\neq k$, taking into account Eq. (\ref{eq:intersection}), respectively.
Then, we allow two types of CP transformations of $\{X^I, {\cal F}_I \}$ and three-form basis to realize Eq. (\ref{eq:CPtrf}):
\begin{itemize}
\item  $\{\alpha_{i}^{({\rm CP})}, \beta^{({\rm CP})\tilde{0}}\} \in  H^{3({\rm CP})}_+({\cal M}),\,\,\{\alpha_{\tilde{0}}^{({\rm CP})}, \beta^{({\rm CP})i}\} \in  H^{3({\rm CP})}_-({\cal M})$ ($i=1,2,...,h^{2,1}$)   
\begin{align}
X^0 &\rightarrow{+\bar{X}^0}, \quad X^i \rightarrow{- \bar{X}^i}, \quad {\cal F}_0 \rightarrow{-\bar{{\cal F}}_0},\quad
    {\cal F}_i \rightarrow{+ \bar{{\cal F}}_i},
    \nonumber\\
    \alpha_{\tilde{0}} &\rightarrow{- \alpha_{\tilde{0}}},\quad
    \alpha_i \rightarrow{+ \alpha_i},\quad
    \beta^{\tilde{0}} \rightarrow{+ \beta^{\tilde{0}}},\quad
    \beta^i \rightarrow{- \beta^i},
\label{eq:Fcptrf1}
    \end{align}
\item  $\{\alpha_{\tilde{i}}^{({\rm CP})}, \beta^{({\rm CP})0}\} \in  H^{3({\rm CP})}_-({\cal M}),\,\,\{\alpha_{0}^{({\rm CP})}, \beta^{({\rm CP})\tilde{i}}\} \in  H^{3({\rm CP})}_+({\cal M})$ ($\tilde{i}=1,2,...,h^{2,1}$)   
\begin{align}
X^0 &\rightarrow{-\bar{X}^0}, \quad X^i \rightarrow{+\bar{X}^{\tilde{i}}}, \quad {\cal F}_0 \rightarrow{+\bar{{\cal F}}_0},\quad
    {\cal F}_{\tilde{i}} \rightarrow{- \bar{{\cal F}}_{\tilde{i}}},
    \nonumber\\
    \alpha_0 &\rightarrow{+ \alpha_0},\quad
    \alpha_{\tilde{i}} \rightarrow{-\alpha_{\tilde{i}}},\quad
    \beta^0 \rightarrow{- \beta^0},\quad
    \beta^{\tilde{i}} \rightarrow{+ \beta^{\tilde{i}}}.
\label{eq:Fcptrf2}
    \end{align}
\end{itemize}
Recall that we restrict ourselves to the large complex structure regime.

However, it is difficult to achieve these CP transformations for a generic form of the prepotential. 
In the large complex structure regime, the prepotential ${\cal F}(X)= (X^0)^2F(u)$ is indeed expanded as\footnote{We still call $F(u)$ as the prepotential throughout this paper.}
\begin{align}
F(u)  = \frac{1}{3!}\kappa_{ijk}u^iu^ju^k
    +\frac{1}{2!}\kappa_{ij}u^iu^j +\kappa_iu^i +\frac{1}{2}\kappa_0, 
\label{eq:Fdef}
\end{align}
up to geometrical corrections \cite{Hosono:1994ax}.
Here, the coefficients $\kappa_{ijk}$, $\kappa_{ij}$, $\kappa_i$ $\kappa_0$ are the topological quantities determined by the CY data, 
i.e.
\begin{align}
    \kappa_{ijk}&=\int_{\tilde{M}} J_i \wedge J_j \wedge J_k,\quad
    \kappa_{ij}=\frac{1}{2}\int_{\tilde{M}} J_i \wedge J_j^2,
\quad
    \kappa_i = -\frac{1}{24}\int_{\tilde{M}} c_2(\tilde{M}) \wedge J_i,\quad
    \kappa_0 = -\frac{\zeta(3)\chi({\tilde M})}{(2\pi i)^3}.
\end{align}
These classical topological quantities are calculated on the mirror CY threefold $\tilde{M}$, where 
the $(1,1)$-forms are denoted by $J_i$, and the second Chern class and the Euler characteristic of ${\tilde M}$ 
are represented by $c_2(\tilde{M})$ and $\chi({\tilde M})$, respectively.\footnote{Note that the quadratic and linear terms of the prepotential, $\kappa_{ij}$ and $\kappa_i$, are only determined modulo integers, where we adopt the convention in Ref.~\cite{Mayr:2000as}.}

To satisfy the CP transformations (\ref{eq:Fcptrf1}) and (\ref{eq:Fcptrf2}) taking into account $u^i\rightarrow -\bar{u}^i$, 
the prepotential is restricted to be a cubic type, i.e.\footnote{Here and in what follows, we use the same symbol $i$ for the CP-even/odd bases.}
\begin{align}
    F_{\rm cubic} = \frac{1}{3!}\kappa_{ijk}u^iu^ju^k.
\label{eq:Fcubic}
\end{align}
The linear term in the prepotential is also allowed under the CP transformation, but we focus on a strict large complex structure regime of generic CY threefolds in 
the following analysis.

Before analyzing the CP-invariant effective potential, 
we discuss the relation between the CP symmetry and the symplectic modular symmetry of CY threefolds, described by the special geometry. 
Given the holomorphic three-form $\Omega$ expanded on the basis of the symplectic basis $(\alpha_I, \beta^I)$ in $H^3({\cal M},\mathbb{Z})$ as in Eq.~(\ref{eq:Omegaexpand}), 
the symplectic basis transforms under the symplectic modular symmetry as
\begin{align}
    \begin{pmatrix}
       \alpha\\
       \beta\\
    \end{pmatrix}
\rightarrow
    \begin{pmatrix}
       a & b\\
       c & d\\
    \end{pmatrix}
    \begin{pmatrix}
       \alpha\\
       \beta\\
    \end{pmatrix},
\end{align}
with
\begin{align}
        \begin{pmatrix}
       a & b\\
       c & d\\
    \end{pmatrix}
    \in Sp(2h^{2,1}+2;\mathbb{Z}).
\end{align}
That corresponds to the group of $(2h^{2,1}+2)\times (2h^{2,1}+2)$ matrices preserving the symplectic matrix:
\begin{align}
    \Sigma = 
    \begin{pmatrix}
       0 & \bf{1} \\
       -\bf{1} & 0\\
    \end{pmatrix}
.
\end{align}

Under the symplectic transformations, the so-called 
period integrals of the holomorphic three-form $\Omega$ on the three-cycles $\{A^I, B_I\}$ transform as
\begin{align}
    \Pi =
    \begin{pmatrix}
       \int_{A}\Omega \\
       \int_{B}\Omega\\
    \end{pmatrix}
    =
    \begin{pmatrix}
       \Vec{X}\\
       \Vec{{\cal F}}\\
    \end{pmatrix}
\rightarrow
        \begin{pmatrix}
       d & c\\
       b & a\\
    \end{pmatrix}
    \begin{pmatrix}
       \Vec{X}\\
       \Vec{{\cal F}}\\
    \end{pmatrix}
,
    \label{eq:Pitrf}
\end{align}
with $\vec{X}=\{X^I\}$ and $\vec{{\cal F}}=\{ {\cal F}_I\}$. 
These transformations are reflected by the fact that holomorphic three-form itself is 
invariant under the symplectic transformation.
Let us introduce the period matrix
\begin{align}
    {\cal F}_{IJ} = \frac{\partial^2}{\partial u^I \partial u^J}{\cal F} = \partial_I {\cal F}_J,
\end{align}
which transforms under the symplectic transformation as
\begin{align}
    {\cal F}_{IJ}\rightarrow \left((a{\cal F}+b)(c{\cal F}+d)^{-1}\right)_{IJ},
\end{align}
analogous to the period matrices of Riemann surfaces with genus $g$. 
Correspondingly, when we take the gauge $X^0=1$, the complex structure 
moduli form a vector-valued modular form of $Sp(2h^{2,1}+2;\mathbb{Z})$, 
\begin{align}
    u \rightarrow (c{\cal F}+d)u,
\end{align}
with $u^0=X^0=1$. 
We recall that the K\"ahler potential
\begin{align}
    K_{\rm cs} = -\ln \biggl[ -i \Pi^\dagger \cdot \Sigma \cdot \Pi\biggl],
\end{align}
is indeed invariant under the symplectic transformation of the period integrals (\ref{eq:Pitrf}). 
Such a symplectic modular symmetry arises from the fact that 
the CY moduli space is described by the special geometry.

On the toroidal background $T^{2n}$, the geometrical space group is the $SL(2n,\mathbb{Z})$ 
modular group.
For example, the CP transformation on $T^2$ is identified with the outer automorphism 
of $SL(2, \mathbb{Z})$~\cite{Nilles:2018wex,Novichkov:2019sqv}. 
It is interesting to ask whether CP is embedded into the symplectic modular transformation on CY threefolds. 
The CP transformation we discussed corresponds to the 
following transformation for the period integrals,
\begin{align}
    \Pi = 
    \begin{pmatrix}
       X^0\\
       X^i\\
       {\cal F}_0\\
       {\cal F}_i\\
    \end{pmatrix}
    \rightarrow
     \pm
     \begin{pmatrix}
       X^0\\
       -\bar{X}^i\\
       -\bar{{\cal F}}_0\\
       \bar{{\cal F}}_i\\
    \end{pmatrix}   
      = \pm
     \begin{pmatrix}
       1 & 0 & 0 & 0\\
       0 & -\bf{1} & 0 & 0\\
       0 & 0 & -1 & 0\\
       0 & 0 & 0 & \bf{1}\\
    \end{pmatrix}
    \bar{\Pi}
    .
\end{align}
We find that the matrix
\begin{align}
{\cal CP}=\pm
     \begin{pmatrix}
       1 & 0 & 0 & 0\\
       0 & -\bf{1} & 0 & 0\\
       0 & 0 & -1 & 0\\
       0 & 0 & 0 & \bf{1}\\
    \end{pmatrix}
    \notin Sp(2h^{2,1}+2; \mathbb{Z}),
\label{eq:CPdef}
\end{align}
is {\it not} the element of $Sp(2h^{2,1}+2; \mathbb{Z})$, 
due to the property ${\cal CP}^T
 \cdot \Sigma \cdot 
{\cal CP}
=\Sigma^T \neq \Sigma$. 
We have two options to define the CP transformation as in Eq.~(\ref{eq:CPdef}), but 
both of them share the common properties. 
Hence, we consider the case with the positive sign in Eq.~(\ref{eq:CPdef}) without loss 
of generality. 
Under CP and symplectic modular transformation $\gamma$, we obtain
\begin{align}
    \Pi \xrightarrow{\rm CP} {\cal CP}\bar{\Pi} 
    \xrightarrow{\gamma} {\cal CP}\cdot \gamma \bar{\Pi} 
    \xrightarrow{{\rm CP}^{-1}} {\cal CP}\cdot \gamma \cdot {\cal CP}^{-1} \Pi,
\end{align}
namely
\begin{align}
   {\cal CP}\cdot \gamma \cdot {\cal CP}^{-1}
   =
   \left(
       \begin{array}{cc}
       \hat{\sigma}^3 & 0\\
       0 & -\hat{\sigma}^3
    \end{array}
    \right)
   \left(
       \begin{array}{cc}
       d & c\\
       b & a
    \end{array}
    \right)
   \left(
       \begin{array}{cc}
       \hat{\sigma}^3 & 0\\
       0 & -\hat{\sigma}^3
    \end{array}
    \right)
    =
    \begin{pmatrix}
       \hat{\sigma}^3 d\hat{\sigma}^3 & -\hat{\sigma}^3 c\hat{\sigma}^3\\
       -\hat{\sigma}^3 b\hat{\sigma}^3 & \hat{\sigma}^3 a\hat{\sigma}^3\\
    \end{pmatrix}
    =
    \begin{pmatrix}
       d & -c\\
       -b & a\\
    \end{pmatrix},
\end{align}
with
\begin{align}
    \hat{\sigma}^3 \equiv 
    \begin{pmatrix}
       1 & 0\\
       0 & -\bf{1}\\
    \end{pmatrix}
    .
\end{align}

In this way, we can define the outer automorphism ${\cal Q}$:
\begin{align}
    \gamma =
     \begin{pmatrix}
       d & c\\
       b & a\\
    \end{pmatrix}  
    \rightarrow
    {\cal Q}(\gamma)\equiv
       {\cal CP}\cdot \gamma \cdot {\cal CP}^{-1}
           =
    \begin{pmatrix}
       d & -c\\
       -b & a\\
    \end{pmatrix}
    ,
\end{align}
satisfying
\begin{align}
    {\cal Q}(\gamma_1){\cal Q}(\gamma_2)=   {\cal CP}\cdot \gamma_1 \cdot {\cal CP}^{-1}{\cal CP}\cdot \gamma_2 \cdot {\cal CP}^{-1}
    ={\cal Q}(\gamma_1\gamma_2),
\end{align}
and no group element $\hat{\gamma}\in Sp(2h^{2,1}+2, \mathbb{Z})$ exists such that ${\cal Q}(\gamma)=\hat{\gamma}^{-1}\gamma\hat{\gamma}$.
In this respect, we argue that CP has a geometrical origin in the CY moduli space, 
namely the outer automorphism of the $Sp(2h^{2,1}+2, \mathbb{Z})$ modular group. 
The whole group is described by the semidirect product group $Sp(2h^{2,1}+2, \mathbb{Z}) \rtimes {\cal CP}$, since there exists a group homomorphism from ${\cal CP}$ to the automorphism group of $Sp(2h^{2,1}+2, \mathbb{Z})$ and $Sp(2h^{2,1}+2, \mathbb{Z})$ is a normal subgroup of $Sp(2h^{2,1}+2, \mathbb{Z}) \rtimes {\cal CP}$. 
That is a natural extension of the $SL(2,\mathbb{Z})\simeq Sp(2, \mathbb{Z})$ modular 
group known in the $T^2$ toroidal background \cite{Nilles:2018wex,Novichkov:2019sqv}, 
where the whole group is given by the “generalized modular group" 
$GL(2,\mathbb{Z})\simeq SL(2, \mathbb{Z}) \rtimes {\cal CP}$.\footnote{We can treat an element of $Sp(2 h^{2,1}+2, \mathbb{Z}) \rtimes {\cal CP}$ as a pair $(g, h)$ where $g \in Sp(2 h^{2,1}+2, \mathbb{Z})$, $h \in {\cal CP} \simeq \mathbb{Z}_2$, and a product between two pairs is represented as $(g_1, h_1) (g_2, h_2 ) = (g_1 h_1^{-1} g_2 h_1, h_1 h_2)$ with respect to the outer automorphism. Nevertheless, in the toroidal case, it is enough to consider a set of elements $(g, h)  (h \neq 1)$ as a negative determinant part of $GL(2, \mathbb{Z})$, so we simply use $GL(2, \mathbb{Z})$ instead of $SL(2, \mathbb{Z}) \rtimes {\cal CP}$ thanks to the isomorphism. However, because of the explicit form of $\hat{\sigma}^3$, we cannot think $Sp(2 h^{2,1}+2, \mathbb{Z}) \rtimes {\cal CP}$ as the same with the toroidal case for odd $h^{2,1}$, where ${\cal CP}$ has a positive determinant. For even $h^{2,1}$, ${\cal CP}$ has a negative determinant then it could be similar to the toroidal case that we can distinguish new elements of the whole group $Sp(2 h^{2,1}+2, \mathbb{Z}) \rtimes {\cal CP}$ that are not in original $Sp(2 h^{2,1} +2, \mathbb{Z})$ by their signs of determinants.}

\section{CP-invariant flux compactifications and discrete symmetry}
\label{sec:3}

In this section, we introduce the fluxes into the previous effective action 
to determine the size of CP-violation, namely the vacuum expectation values of moduli fields. 
We mainly focus on the effective action of Type IIB flux compactifications on CY orientifolds with $O3/O7$-planes, but it is straightforward to extend our analysis to T-dual Type IIA flux 
compactifications as well as the heterotic string theory in the large volume/complex structure and weak coupling regime. 
After examining the relation between CP and orientifold actions in section \ref{sec:3_1}, 
we first derive the CP-invariant flux compactifications and next discuss the 
discrete $\mathbb{Z}_2$ or $\mathbb{Z}_4$ symmetry appearing in the effective action in section \ref{sec:3_2}.

\subsection{CP and orientifold actions}
\label{sec:3_1}

Before studying the relation between the CP transformation and orientifold projection in the context of Type II string theory, we remind the CP transformation of the K\"ahler form in the context of heterotic string theory. 
The orientation reversing in the 6D space is characterized by the transformation of 
the holomorphic three-form $\Omega$ and the K\"ahler form $J$:
\begin{align}
    \Omega \rightarrow -\bar{\Omega},
    \quad
    J \rightarrow -J.
    \label{eq:CPJ}
\end{align}
Since we examined the CP transformation of the holomorphic three-form in the previous section, we focus on the CP transformation of the K\"ahler form on CY threefolds ${\cal M}$ in more detail. 

The Kalb-Ramond two-form ($B$) and the K\"ahler form ($J$) are expanded 
by the basis of $H^{1,1}({\cal M},\mathbb{R})$,
\begin{align}
    B = \sum_{a=1}^{h^{1,1}} b^a(x) w_a,\quad 
    J = \sum_{a=1}^{h^{1,1}} t^a(x) w_a,
\end{align}
with $h^{1,1}$ being the dimension of  $H^{1,1}({\cal M},\mathbb{R})$. 
Here, 4D K\"ahler axions $b^a$ and volume moduli associated with two-cycles $t^a$ (saxions) are 4D CP-odd and -even fields, respectively. 
Recalling that the space of harmonic forms splits under the CP transformation into even and odd eigenspaces, $H^2({\cal M})=H^{2({\rm CP})}_+({\cal M})\oplus H^{2({\rm CP})}_-({\cal M})$,
the orientation reversing of the K\"ahler form (\ref{eq:CPJ}) requires that the bases $w_a$ are CP-odd bases, i.e.,
\begin{align}
    w_l \in  H^{1,1({\rm CP})}_-({\cal M}),
    \label{eq:wCPodd}
\end{align}
where $l=1,2,...,h^{1,1}_-|_{\rm CP}$ with $h^{1,1}_-|_{\rm CP}$ being the dimension of $H^{1,1({\rm CP})}_-({\cal M})$, transforming as $w_l \rightarrow - w_l$. 
Thus, the simultaneous transformation of 4D parity and 6D orientation reversing (i.e., CP transformation) of complexified K\"ahler form $J_c=B+iJ$ is provided by the anti-holomorphic transformation:
\begin{align}
    J_c \rightarrow \bar{J}_c.
\end{align}

Let us move to Type IIA/IIB CY orientifolds, where a part of complex structure moduli and K\"ahler moduli also determines the strong CP phase and Cabbibo-Kobayashi-Maskawa phase. (See for more details, e.g., Ref.\cite{Blumenhagen:2006ci}.)
It indicates that the 4D CP transformation is understood from the 10D point of view, i.e., 
simultaneous transformations of 4D parity and 6D orientation reversing, as discussed before. 
In subsequent discussions, we examine the relation between the CP transformation and orientifold projections, with an emphasis on Type IIA orientifolds with O6-planes and Type IIB orientifolds with O3/O7-planes whose geometric $\mathbb{Z}_2$ action is represented by ${\cal R}$. (For the details about the orientifold projections, we refer to Refs. \cite{Grimm:2004ua,Grimm:2004uq}.)

\begin{enumerate}
\item Type IIA orientifolds with O6-planes : ${\cal R}\Omega = e^{2i \theta} \bar{\Omega}$,\quad ${\cal R}J = -J$

The constant phase $\theta$ can be absorbed into the definition of holomorphic three-form. 
In particular, $\theta =\pi/2$ case with $z_i\rightarrow -\bar{z}_i$ is the same with the CP transformation in our notation. 
As stated before, the phase of CP transformation $\Omega \rightarrow \pm \bar{\Omega}$ is a matter of conventions. 
Indeed, $\Omega \rightarrow \bar{\Omega}$ is achieved under the coordinate transformation $z_i\rightarrow \bar{z}_i$. 
In this way, the orientifold action is consistent with the 6D orientation reversing by identifying orientifold-even (-odd) basis with the CP-even (-odd) basis in the space of harmonic forms.

\item Type IIB orientifolds with O3/O7-planes : ${\cal R}\Omega = -\Omega$,\quad ${\cal R}J = J$

We first analyze the orientifold action of the holomorphic three-form, expanded in the orientifold-odd bases of $H^{2,1}_-({\cal M},\mathbb{C})$ i.e., $\{\alpha_{\kappa}^{({\rm O})},\beta^{({\rm O})\kappa}\}$,
\begin{align}
    \Omega = X^\kappa \alpha_{\kappa}^{({\rm O})} - F_\kappa \beta^{({\rm O})\kappa},
\end{align}
where $\kappa=0,1,...,h^{2,1}_-$ with $h^{2,1}_-$ being the hodge number. 
In the local patch, orientifold-odd bases $\{\alpha_\kappa\}$ are spanned by $\{dx_i \wedge dx_j \wedge dx_k, dx_i \wedge dx_j \wedge dy_k, dx_i \wedge dy_j \wedge dy_k,dy_i \wedge dy_j \wedge dy_k\}$ with $i\neq j\neq k$, 
satisfying $\{\alpha_\kappa^{({\rm O})}\} \rightarrow -\{\alpha_\kappa^{({\rm O})}\}$ under $z_i\rightarrow -z_i$ ($x_i,y_i\rightarrow -x_i,-y_i$). The CP-odd and -even bases classified by $z_i\rightarrow -\bar{z}_i$ ($x_i\rightarrow -x_i, y_i\rightarrow y_i$) are also spanned by a part of the orientifold-odd bases, from the fact that their bases in the local coordinates are given by $\{dx_i \wedge dx_j \wedge dx_k, dx_i \wedge dy_j \wedge dy_k\}$ and $\{dx_i \wedge dx_j \wedge dy_k, dy_i \wedge dy_j \wedge dy_k\}$ with $i\neq j\neq k$, respectively. Hence, the CP invariant action is consistent with the orientifold action, since CP-odd/even bases are more restrictive than the orientifold-odd bases. 
In this way, the form of the prepotential is restricted to be a cubic type, although the generic form is allowed in Type IIB CY orientifolds with O3/O7-planes.

Next, we examine the orientifold action with an emphasis on K\"ahler form $J$ and Ramond-Ramond four-form $C_4$ consisting of the K\"ahler moduli.\footnote{We omit the contribution from the orientifold-odd moduli, for simplicity.}  
It was known that they are expanded in the orientifold-even basis $\{w_l^{({\rm O})}\}$ of $H^{1,1}_+({\cal M},\mathbb{R})$,
\begin{align}
    J = t^l w_l^{({\rm O})},\quad
    C_4 = \rho^l \tilde{w}_l^{({\rm O})},\quad
\end{align}
where $\{\tilde{w}_l^{({\rm O})}\}$ denote the hodge dual of $\{w_l^{({\rm O})}\}$, $l=1,2,...,h^{1,1}_+$, with $h^{1,1}_+$ being the dimension of $H^{1,1}_+({\cal M},\mathbb{R})$. Note that the orientifold-even K\"ahler moduli are given by the integral of $C_4+iJ\wedge J$ over the four-cycle,
\begin{align}
    T^l = \rho^l + i \kappa_{lmn}t^mt^n,
\end{align}
with $\kappa_{lmn}$ being the triple intersection numbers. 
As analyzed in the context of heterotic string, the CP-transformation of the K\"ahler form requires that the (1,1)-forms are elements of $H^{1,1({\rm CP})}_-({\cal M})$ (\ref{eq:wCPodd}), and both $J$ and $C_4$ are expanded in the same CP-odd bases due to the existence of supersymmetry,
\begin{align}
    J = t^l w_l^{({\rm CP})},\quad
    C_4 = \rho^l \tilde{w}_l^{({\rm CP})}.
\end{align}
It leads to the anti-holomorphic transformation of the K\"ahler moduli 
\begin{align}
    T^l \rightarrow -\bar{T}^l,
\end{align}
since the RR axions $\rho^l$ are 4D CP-odd fields, namely the axions, similar to the Kalb-Ramond B-field. 

Let us check the consistency between the bases split by the orientifold and CP actions. 
Note that the basis of $H^{1,1}_+({\cal M},\mathbb{R})$ is locally given in  $w_l =w_{l}^{i\bar{j}} dz_i \wedge d\bar{z}_{\bar{j}}$ which transforms as $w_l\rightarrow w_l$ under the orientifold action $z_i\rightarrow -z_i$ and $w_l\rightarrow -w_l$ under the CP action $z_i\rightarrow -\bar{z}_i$. 
Thus, the orientifold-even and CP-odd bases are consistent with each other. 
So far, we have focused on the CP transformations of moduli fields relevant to flux compactifications, but it is straightforward to analyze the CP transformation of other supergravity fields in a similar way.

\end{enumerate}

\subsection{Flux compactifications}
\label{sec:3_2}

We study the flux compactification in Type IIB string on CY orientifolds ${\cal M}$.
In addition to the effective action of the complex structure moduli shown in Sec.~\ref{sec:2}, 
the moduli effective action includes the axio-dilaton and the K\"ahler moduli,
\begin{align}
    K &= -\ln (i(\bar{S}-S))-\ln \biggl[-i \int_{{\cal M}} \Omega \wedge \bar{\Omega}\biggl] -2\ln {\cal V},
    \nonumber\\
    W &= \int_{{\cal M}} G_3 \wedge \Omega,
\label{eq:KW}
\end{align}
where ${\cal V}$ represents for the CY volume in the Einstein frame, depending on the K\"ahler moduli $T^l$ of ${\cal M}$. 
Here, we introduce the flux-induced superpotential \cite{Gukov:1999ya} generated by 
a linear combination of Ramond-Ramond (RR) three-form flux $F_3$ and Neveu-Schwarz (NS) three-form flux $H_3$, 
namely $G_3=F_3 -S H_3$ as a function of the axio-dilaton $S$. 
Since the CY volume is invariant under the orientation reversing transformation, 
kinetic terms of the K\"ahler moduli as well as the complex structure moduli are CP-invariant quantities. 
This is because both of them originate from the Einstein-Hilbert 
action in 10D supergravity action, which is invariant under the 10D proper Lorentz transformation. 
Since the axio-dilaton $S$ is a 4D CP-odd field,
\begin{align}
    S \rightarrow{-\bar{S}},
\label{eq:Strf}
\end{align}
the kinetic term of $S$ is also invariant under the CP transformation. 
However, the three-form fluxes generically break CP as well as the 6D Lorentz symmetry. 
Hence, not all the flux quanta are allowed in the CP-invariant effective action. 
We classify the possible pattern of 
CP-invariant three-form fluxes inserted on three-cycles of CY threefolds.

The CP invariance of the 4D effective action requires $W \rightarrow{e^{i\gamma}\bar{W}}$ for the superpotential, where $\gamma$ denotes a complex phase. 
Taking into account the CP transformation of $\Omega$ in Eq. (\ref{eq:Omegatrf}), 
the three-form flux $G_3$ transforms into
\begin{align}
    G_3 &\rightarrow{-e^{i\gamma}\bar{G}_3},
\end{align}
from which the integral flux quanta require $\gamma=0$ or $\pi$.

Recalling that the axio-dilaton $S$ is a 4D CP-odd field, 
RR and NS three-form fluxes should transform as
\begin{align}
   \left\{
\begin{array}{l}
   \gamma=0,\quad
    F_3 \rightarrow{-F_3}, \quad H_3 \rightarrow{H_3}
   \\
   \gamma=\pi,\quad
    F_3 \rightarrow{F_3}, \quad H_3 \rightarrow{-H_3}
    \end{array}
\right.
.
\end{align}
Hence, the expansion of the three-form fluxes on the symplectic basis is categorized into two classes:
\begin{itemize}
    \item $\gamma =0$
\begin{align}
    F_3 &= f^0 \alpha_0 + f_i \beta^i,
    \nonumber\\
    H_3 &= h^i \alpha_i +h_0 \beta^0.
\end{align}
    \item $\gamma=\pi$
\begin{align}
    F_3 &= f^i \alpha_i + f_0 \beta^0,
    \nonumber\\
    H_3 &= h^0 \alpha_0 +h_i \beta^i.
\end{align}
\end{itemize}

As a result, we obtain two classes of 4D CP-invariant effective action in the large complex structure 
regime of CY threefolds. 
In both classes, 
the K\"ahler potential is described by
\begin{align}
    K_{\rm cs} &= -\ln \biggl[-i \int_{{\cal M}} \Omega \wedge \bar{\Omega}\biggl]
               = -\ln \biggl[-i (\bar{u}^IF_I -u^I\bar{F}_I) \biggl]
     \nonumber\\
    &= -\ln \biggl[\frac{i}{3!}\kappa_{ijk}(u^i-\bar{u}^i)(u^j-\bar{u}^j)(u^k-\bar{u}^k) \biggl],
\label{eq:Kcs}
\end{align}
where we employ Eq.~(\ref{eq:Fcubic})\footnote{Here and in what follows, we adopt the 
cubic-type prepotential under the gauge $X^0=1$, and the dimension of the complex structure moduli space is given by $h^{2,1}_-$.} with $i=1,2,\cdots, h^{2,1}_-$.
However, for the superpotential we have two options:
\begin{itemize}
    \item $\gamma = 0$
    \begin{align}
        W &= -f_iu^i -f^0(2F-u^i \partial_i F)  -S\left( -h_0  -h^i\partial_iF\right)
        \nonumber\\
          &= -f_iu^i -f^0\left(-\frac{1}{6}\kappa_{ijk}u^iu^ju^k \right)
             +h_0S +h^iS \left( \frac{1}{2}\kappa_{ijk}u^ju^k \right).
\label{eq:Wodd}
    \end{align}
    \item $\gamma = \pi$
    \begin{align}
        W &=  -f_0 -f^i\partial_i F -S\biggl[-h_iu^i  -h^0(2F-u^i \partial_i F)\biggl]
        \nonumber\\
          &= -f_0 -f^i \left( \frac{1}{2}\kappa_{ijk}u^ju^k\right) 
            +h_iu^iS +h^0S\left(-\frac{1}{6}\kappa_{ijk}u^iu^ju^k \right).
\label{eq:Weven}
    \end{align}
\end{itemize}
Hence, the CP-invariant superpotential is restricted to be either odd 
or even polynomials with respect to the moduli fields. 
Note that these fluxes induce the D3-brane charge:
\begin{align}
   N_{\rm flux}= \int H_3 \wedge F_3 
    =
            \left\{
        \begin{array}{l}
              -f^0h_0 + \sum_i f_ih^i,\quad (\gamma=0)
              \\
              f_0 h^0 -\sum_i f^ih_i,\quad (\gamma=\pi)
        \end{array}
        \right.
        ,
\end{align}
which should be canceled by mobile D3-branes and orientifold 
contributions.

Finally, we comment on the (accidental) symmetry which arises in the CP-invariant effective action 
from the field theoretical point of view. 
The superpotential of even degree with respect to the moduli fields in Eq.~(\ref{eq:Weven}) 
is invariant under the discrete $\mathbb{Z}_2$ symmetry (not related to the discrete R-symmetry)
    \begin{align}
u^i \rightarrow -u^i,\quad S\rightarrow -S.
    \end{align}
For the superpotential with the odd degree of moduli fields, 
one can assign the R-charge 2 for all the moduli fields due to the existence of linear terms. However, the presence of cubic terms breaks the continuous R-symmetry 
and it results in the discrete $\mathbb{Z}_4$ R-symmetry in the above superpotential. 
Note that in the superpotential of both odd and even degrees, the K\"ahler potential is also invariant under 
the discrete  $\mathbb{Z}_2$ and $\mathbb{Z}_4$ symmetries, taking into account transformations of 
the axio-dilaton and the complex structure moduli at the same time.  

Let us consider the generic flux-induced superpotential having both odd and even degrees of polynomials with 
respect to the moduli fields:
    \begin{align}
W= -f_iu^i -f^0(2F-u^i \partial_i F)  -S\left( -h_0  -h^i\partial_i F\right)
 -f_0 -f^i\partial_i F -S\biggl[-h_iu^i  -h^0(2F-u^i \partial_i F)\biggl].
\label{eq:Wgeneric}
    \end{align}
When we impose the discrete $\mathbb{Z}_2$ or $\mathbb{Z}_4$ symmetry for moduli fields 
in the supersymmetric effective action, 
the generic flux-induced superpotential (\ref{eq:Wgeneric}) in the large complex structure 
regime has a CP-invariance 
categorized by two classes, i.e. Eqs.~(\ref{eq:Wodd}) and (\ref{eq:Weven}). 
Hence, the necessary and sufficient condition to possess CP in the moduli effective action 
is the existence of discrete $\mathbb{Z}_2$ or $\mathbb{Z}_4$ symmetry and the supersymmetry (SUSY). 
Recalling that the spontaneous CP violation is realized by the non-zero vacuum expectation values 
of axionic fields ${\rm Re}u^i$ and ${\rm Re}S$, 
the discrete $\mathbb{Z}_2$ or $\mathbb{Z}_4$ symmetry is spontaneously broken at the CP-breaking vacua.

\section{CP-conserving and -breaking vacua}
\label{sec:4}

We are ready to analyze the vacuum structure of CP-invariant 
effective action. 
We first discuss the existence of CP-conserving vacua in 
the large complex structure regime of generic CY threefolds in Sec.~\ref{subsec:4_1}. 
Next, we deal with concrete CY threefolds in the large complex structure regime to clarify whether the 
spontaneous CP violation is realized or not. 
It turns out that for the prepotential having a similar structure of factorizable tori in Secs.~\ref{subsec:4_2} and \ref{subsec:4_3}, 
it is difficult to achieve the spontaneous CP violation. 
When the structure of prepotential deviates from the toroidal one,  
one can find the CP-breaking vacua as discussed in detail in Sec.~\ref{subsec:4_4}. Here, we choose a specific CY such as the degree 18 hypersurface in a weighted projective space $\mathbb{CP}_{11169}$. 

\subsection{CP-conserving vacua in generic CY}
\label{subsec:4_1}

Before analyzing the CP-breaking minimum of the 
scalar potential, we investigate the CP-conserving minima satisfying 
the SUSY condition,
\begin{align}
D_IW = \left(\partial_I + K_I\right) W=0,
\end{align}
where $I=\{S, u^i \}$ runs over the axio-dilaton $S$ and all the complex structure moduli $u^i$, 
and $K_I=\partial_I K$. 
Since the effective action in the large volume regime possesses the no-scale structure for the 
K\"ahler moduli, we focus on the dynamics of the axio-dilaton and the complex structure moduli 
in the following analysis.

The superpotential and its covariant derivatives transform under the CP transformations $\{S\rightarrow -\bar{S}, u^i \rightarrow -\bar{u}^i\}$ 
as
\begin{itemize}
    \item $\gamma =0$
\begin{align}
    W &\rightarrow \overline{W},\quad D_SW \rightarrow \overline{D_S W},\quad
    D_i W \rightarrow{\overline{D_iW}},
\end{align}
    \item $\gamma=\pi$
\begin{align}
    W &\rightarrow -\overline{W},\quad D_SW \rightarrow -\overline{D_S W},\quad
    D_i W \rightarrow -{\overline{D_iW}},
\end{align}
\end{itemize}
where we use $K_{\bar{i}}=-K_i$ and $K_i$ is a CP-even function due to the axionic 
shift symmetries. 
Hence, for both $\gamma=0$ and $\pi$ cases, the SUSY minimum $D_IW=0$ has a symmetry, 
$\langle {\rm Re}S\rangle \rightarrow{-\langle {\rm Re}S\rangle}$ and 
$\langle {\rm Re} u^i\rangle \rightarrow{-\langle {\rm Re} u^i\rangle}$. 
It suggests that CP-invariant moduli effective action may have the CP-conserving 
vacua in generic CY flux compactifications. 
Indeed, among the $h^{2,1}_- +1$ number of SUSY conditions $D_I W=0$, 
the CP-conserving vacuum ${\rm Re}S={\rm Re}u^i=0$ is a solution for half of them, namely ${\rm Re}(D_IW)=0$ and ${\rm Im}(D_IW)=0$ for $\gamma = \pi$ and $\gamma=0$, respectively. 
By introducing 
\begin{align}
&{\cal X}^i\equiv \kappa_{ijk}{\rm Re}u^j{\rm Im}u^k,\quad {\cal Y}^i\equiv \kappa_{ijk}{\rm Re}u^j{\rm Re}u^k,\quad {\cal Z}^i\equiv \kappa_{ijk}{\rm Im}u^j{\rm Im}u^k,
\end{align}
one can explicitly check that ${\rm Re}S={\rm Re}u^i=0$ lead to
\begin{itemize}
    \item $\gamma =0$
\begin{align}
    {\rm Im}(D_SW)&= h^i{\cal X}^i +{\rm Im}K_S{\rm Re}W=0,
    \nonumber\\
    {\rm Im}(D_iW)&= f^0{\cal X}^i+{\rm Re}S\kappa_{ijk}h^j{\rm Im}u^k+{\rm Im}S\kappa_{ijk}h^j{\rm Re}u^k +{\rm Im}K_i {\rm Re}W=0,
\label{eq:SUSYeqsodd}
\end{align}
with 
\begin{align}
{\rm Re}W=-f_i{\rm Re}u^i +\frac{f^0}{6}\left({\cal Y}^i -3{\cal X}^i\right){\rm Re}u^i +h_0{\rm Re}S +\frac{h^i}{2}{\rm Re}S\left({\cal Y}^i -{\cal Z}^i\right)-{\rm Im}Sh^i{\cal X}^i.
\end{align}
    \item $\gamma =\pi$
\begin{align}
    {\rm Re}(D_SW)&= h_i{\rm Re}u^i 
    -\frac{h^0}{6}\left({\cal Y}^i  -3{\cal Z}^i\right){\rm Re}u^i -{\rm Im}K_S{\rm Im}W=0,
    \nonumber\\
    {\rm Re}(D_iW)&= -\kappa_{ijk}f^j{\rm Re}u^k 
   +{\rm Re}S\left(h_i -\frac{h^0}{2}{\cal Y}^i +\frac{h^0}{2}{\cal Z}^i\right)+h^0{\rm Im}S{\cal X}^i -{\rm Im}K_i {\rm Im}W=0, 
\label{eq:SUSYeqseven}
\end{align}
with 
\begin{align}
{\rm Im}W=-f^i{\cal X}^i +{\rm Re}S\left(h_i -\frac{h^0}{2}{\cal Y}^i+\frac{h^0}{6}{\cal Z}^i \right){\rm Im}u^i +{\rm Im}S\left(h_i -\frac{h^0}{6}{\cal Y}^i +\frac{h^0}{2}{\cal Z}^i\right){\rm Re}u^i.
\end{align}
\end{itemize}
Note that ${\rm Im}K_I$ are functions of imaginary parts of moduli fields. 

In this way, half of the SUSY conditions $D_IW=0$ are satisfied by ${\rm Re}S={\rm Re}u^i=0$, and the imaginary parts are determined by the remaining SUSY conditions for both $\gamma=0$ and $\gamma=\pi$ cases. 
In the following, we explicitly check the existence of CP-conserving vacua and 
search for the CP-breaking vacua on concrete CY threefolds.

\subsection{One modulus case: $\mathbb{CP}_{11111}[5]$}
\label{subsec:4_2}

We begin with the CY threefold with a single modulus, especially 
the mirror dual of the quintic $\mathbb{CP}_{11111}[5]$ in Ref.~\cite{Candelas:1990rm}, defined by 
the degree 5 hypersurface in a projective space $\mathbb{CP}_{11111}$. 
To realize the CP-invariant moduli potential, we restrict ourselves 
to the large complex structure regime ${\rm Im}U> 1$, 
where we denote the single complex structure modulus by $U$. 
Given the triple intersection number $\kappa_{UUU}=5$, 
the K\"ahler potential is given by
\begin{align}
    K = - \ln(i(\bar{S}-S)) 
        -  \ln \biggl[\frac{5i}{6}(U-\bar{U})^3\biggl],
\end{align}
and the superpotential is categorized by two classes:
    \begin{align}
        W =
        \left\{
        \begin{array}{c}
             -f_U U +\frac{5f^0}{6} U^3
             +h_0S +\frac{5h^U}{2} S U^2
             \quad (\gamma=0)
        \\
             -f_0 -\frac{f^U}{2} U^2 
            +h_U US -\frac{5h^0}{6}SU^3
             \quad (\gamma=\pi)
        \end{array}
        \right.
        ,
    \end{align}
with $\{f^0,f_0, f_U,f^U,h_0,h^0,h^U,h_U\}$ being flux quanta.

By solving SUSY conditions $D_IW =0$ for $S$ and $U$, we find the CP-conserving minima:
\begin{itemize}
    \item $\gamma=0$
\begin{align}
    {\rm Re}U = {\rm Re}S=0,\quad
    {\rm Im}U = \sqrt{\frac{2}{5}}\left(-\frac{3f_Uh_0}{f^0h^U}\right)^{1/4},\quad     {\rm Im}S = \sqrt{\frac{2}{5}}\left(-\frac{f^0}{3h_0}\right)^{1/4}\left(\frac{f_U}{h^U}\right)^{3/4},
\end{align}
    \item $\gamma=\pi$
\begin{align}
    {\rm Re}U = {\rm Re}S=0,\quad
    {\rm Im}U = \sqrt{2}\left(-\frac{3f_0h_U}{5f^Uh^0}\right)^{1/4},\quad     {\rm Im}S = \sqrt{\frac{1}{2}}\left(\frac{3f_0}{5h^0}\right)^{1/4}\left(-\frac{f^U}{h_U}\right)^{3/4},
\end{align}
\end{itemize}
and four classes of degenerate CP-conserving and -breaking minima:
\begin{align}
&{\rm (i)}\,
    |U|^2 =\frac{6f_U}{5f^0}=-\frac{2h_0}{5h^U},\quad
    S = -\frac{f^0}{3h^U}\bar{U},
\nonumber\\
&{\rm (ii)}\,
     |U|^2 =-\frac{2f_U}{5f^0}=\frac{6h_0}{5h^U},\quad
    |S|^2 =\left(\frac{f^0}{h^U} \right)^2|U|^2,\quad
    {\rm Im}S = \frac{5(f^0)^2({\rm Im}U)^3}{8f_Uh^U+15f^0h^U({\rm Im}U)^2}, 
\end{align}
for $\gamma=0$ and 
\begin{align}
&{\rm (iii)}\,
    |U|^2 =\frac{6h_U}{5h^0},\quad
    S = -\frac{f^U}{2h_U}U,\quad
    f_0 = -\frac{3f^Uh_U}{5h^0},
\nonumber\\
&{\rm (iv)}\,
    |U|^2 =-\frac{2h_U}{5h^0},\quad
    |S|^2 =-\frac{(f^U)^2}{10h^0h_U},\quad
    {\rm Im}S = \frac{5f^Uh^0({\rm Im}U)^3}{4(h_U)^2-30h^0h_U({\rm Re}U)^2},\quad
    f_0 = -\frac{f^Uh_U}{15h^0},
\end{align}
for $\gamma=\pi$, respectively. 
Hence, we cannot realize the spontaneous CP-violation in the large complex structure regime. 
This is because the prepotential has a structure similar to one of  the toroidal background 
with the overall complex structure modulus, where it was pointed out in Ref.~\cite{Kobayashi:2020uaj} that 
the spontaneous CP violation is difficult to achieve. 
This argument also holds for other one-parameter CY threefolds in the large complex structure regime by 
changing the value of triple intersection number, for instance 
the mirror dual of $\mathbb{CP}_{11112}[6]$, $\mathbb{CP}_{11114}[8]$ and 
$\mathbb{CP}_{11125}[10]$ defined on a single polynomial in an ambient weighted projective space. 

Interestingly, for a particular choice of fluxes, CP is embedded into $SL(2,\mathbb{Z})_S$ duality group 
of the axio-dilaton and/or the modular symmetry of the complex structure moduli. 
For instance, the vacuum expectation value of the axio-dilaton in the solution (i) 
is given by
\begin{align}
|S|^2=\frac{2f^0f_U}{15(h^U)^2}=1
\label{eq:S=1}
\end{align}
by setting $f^0f_U=15(h^U)^2/2$. 
The $S$-transformation of the $SL(2,\mathbb{Z})_S$ duality group at the vacuum (\ref{eq:S=1})
\begin{align}
S\rightarrow -1/S = -\bar{S}
\end{align}
corresponds to the CP transformation in Eq.~(\ref{eq:Strf}). 
In this respect, CP is unified into the duality group for a particular choice of fluxes.

\subsection{Two moduli case: $\mathbb{CP}_{11222}[8]$}
\label{subsec:4_3}

The next example is the CY threefold with two complex structure moduli 
labelled by $u^1$ and $u^2$. 
In particular, we deal with the mirror dual of CY threefold defined by 
the degree 8 hypersurface in a weighted projective space $\mathbb{CP}_{11222}$ studied 
in Refs.~\cite{Berglund:1993ax,Candelas:1993dm}, 
where the triple intersection numbers are specified by
\begin{align}
    \kappa_{111}=8,\qquad
    \kappa_{112}=4,
\end{align}
and otherwise 0. 
By restricting ourselves to the large complex structure regime $\{ {\rm Im}u^1,{\rm Im}u^2 > 1\}$, 
the K\"ahler potential is given by
\begin{align}
    K &= - \ln(i(\bar{S}-S)) 
        -  \ln \biggl[\frac{i}{6}\left(8(u^1-\bar{u}^1)^3+12(u^1-\bar{u}^1)^2(u^2-\bar{u}^2)\right)\biggl],
\end{align}
and the superpotential is categorized by two classes:
\begin{itemize}
    \item $\gamma=0$
\end{itemize}
\begin{align}
        W &=-f_1 u^1 -f_2 u^2 +\frac{f^0}{6}(u^1)^2 \left(8u^1+12u^2\right)
             +h_0S +\frac{h^1}{2} S \left(8(u^1)^2+8u^1u^2\right)
             +\frac{h^2}{2} S \left(8u^1u^2\right),
\end{align}
\begin{itemize}
        \item $\gamma=\pi$
\end{itemize}
\begin{align}
        W&=-f_0 -\frac{f^1}{2}\left(8(u^1)^2+8u^1u^2\right)
             -\frac{f^2}{2}\left(8u^1u^2\right)
            +(h_1 u^1+h_2u^2)S -\frac{h^0}{6}S\left(8(u^1)^3+12(u^1)^2u^2\right),
    \end{align}
where we denote the flux quanta $\{f^0,f_0, f_{1,2},f^{1,2},h_0,h^0,h^{1,2},h_{1,2}\}$.

By solving SUSY conditions $D_IW=0$ with $I=S,u^1,u^2$, we find CP-conserving 
solutions:
\begin{itemize}
    \item $\gamma=0$
\begin{align}
{\rm Re}u^1&={\rm Re}u^2={\rm Re}S=0,
\nonumber\\
{\rm Im}u^1 &= 2^{-3/4}\left(\frac{3f_2h_0}{f^0(2h^2-h^1)}\right)^{1/4},\quad
    {\rm Im}u^2 = -\frac{h_0}{2f^0}\frac{{\rm Im}S}{({\rm Im}u^1)^{2}} -\frac{2}{3}{\rm Im}u^1,
    \nonumber\\
    {\rm Im}S &=
    2^{-1}\left(\frac{f^0(2f_2-3f_1)}{h_0(h^1+h^2)}\right)^{1/2}\left(\frac{f_2h_0}{6f^0(2h^2-h^1)}\right)^{1/4}.
\end{align}
    \item $\gamma=\pi$
\begin{align}
{\rm Re}u^1&={\rm Re}u^2={\rm Re}S=0,
\nonumber\\
{\rm Im}u^1 &= 2^{-3/4}\left(\frac{3f_0h_2}{h^0(2f^2-f^1)}\right)^{1/4},\quad
    {\rm Im}u^2 = \frac{-3h_1+2h_2}{12(f^1+f^2)}{\rm Im}S -\frac{2}{3}{\rm Im}u^1,
    \nonumber\\
    {\rm Im}S &=
    2\left(\frac{f^1+f^2}{h^0h_2(2h_2-3h_1)}\right)^{1/2}\left(6f_0h^0h_2(2f^2-f^1) \right)^{1/4}.
\end{align}
\end{itemize}
On the other hand, we rely on the numerical search to find the CP-breaking 
 vacua in both $\gamma=0$ and $\pi$ cases.\footnote{Specifically, we used ``FindRoot" of Mathematica (v12.0) to solve SUSY conditions $D_I W =0$ with randomly generated fluxes and initial values of moduli fields.}  
Numerical search for the randomly generated $2\times 10^7$ dataset of fluxes within $-30 \leq \{f^0,f_0, f_{1,2},f^{1,2},h_0,h^0,h^{1,2},h_{1,2}\} \leq 30$ leading to $0\leq N_{\rm flux}\leq 150$ allows only $5.8\times 10^3$ and 87 stable CP-conserving vacua for $\gamma=0$ and $\gamma=\pi$, respectively. 
The reason why the CP-breaking vacuum is absent is that the K\"ahler 
potential and the superpotential have a similar structure with the toroidal 
one due to the torus-type prepotential. 
Indeed, when we redefine the modulus field as $12u^2=-8u^1 +u$, 
the prepotential is given by
\begin{align}
F(u)&= \frac{1}{6}\left( 8(u_1)^3 +12(u_1)^2u_2\right)=\frac{1}{6}(u_1)^2u.
\end{align}
Because of the moduli redefinition, the prepotential has a similar structure to the factorizable $T^6$ torus by identifying the two complex structure moduli with the identical one.

\subsection{CP-breaking vacua on $\mathbb{CP}_{11169}[18]$}
\label{subsec:4_4}

Finally, we analyze the different two-parameter CY threefold, especially 
the mirror dual of the degree 18 hypersurface in a weighted projective space $\mathbb{CP}_{11169}$ 
studied in Ref.~\cite{Candelas:1994hw}, where the complex structure moduli are labelled by $u^1$ and $u^2$. 
We can also consider the original CY threefold as follows.\footnote{For more details, see, e.g. Ref.~\cite{Giryavets:2003vd}.}
Taking into account a $G=\mathbb{Z}_6\times \mathbb{Z}_{18}$ discrete symmetry of this CY, 
the complex structure moduli space parametrized by $u^1$ and $u^2$ is invariant under this action. 
Other non-invariant complex structure moduli can 
be fixed at the fixed points under $G$, thanks 
to the three-form fluxes along the $G$-invariant three-forms. 
Hence, the period integrals in the mirror dual of the CY 
are the same with the original two-parameter CY threefold. 

To realize the CP-invariant moduli potential, we further restrict ourselves 
to the large complex structure regime $\{{\rm Im}u^1,{\rm Im}u^2 > 1\}$. 
Given the non-vanishing triple intersection numbers $\kappa_{111}=9$, $\kappa_{112}=3$ and $\kappa_{122}=1$, 
the K\"ahler potential is given by
\begin{align}
    K = - \ln(i(\bar{S}-S)) 
        -  \ln \biggl[\frac{i}{6}\left(9(u^1-\bar{u}^1)^3+9(u^1-\bar{u}^1)^2(u^2-\bar{u}^2)+3(u^1-\bar{u}^1)(u^2-\bar{u}^2)^2\right)\biggl],
\end{align}
and the superpotential is categorized by two classes:
    \begin{align}
        W =
        \left\{
        \begin{array}{l}
             -f_1 u^1 -f_2 u^2 +\frac{f^0}{6} \left(9(u^1)^3+9(u^1)^2u^2+3u^1(u^2)^2\right)
             \\
             +h_0S +\frac{h^1}{2} S \left(9(u^1)^2+6u^1u^2+(u^2)^2\right)
             +\frac{h^2}{2} S \left(3(u^1)^2+2u^1u^2\right)
             \quad (\gamma=0)
        \\
        \\
             -f_0 -\frac{f^1}{2}\left(9(u^1)^2+6u^1u^2+(u^2)^2\right)
             -\frac{f^2}{2}\left(3(u^1)^2+2u^1u^2\right)
            \\
            +(h_1 u^1+h_2u^2)S -\frac{h^0}{6}S\left(9(u^1)^3+9(u^1)^2u^2+3u^1(u^2)^2\right)
             \quad (\gamma=\pi)
        \end{array}
        \right.
        ,
    \end{align}
where we denote the flux quanta $\{f^0,f_0, f_{1,2},f^{1,2},h_0,h^0,h^{1,2},h_{1,2}\}$. These integral flux quanta are constrained by the tadpole cancellation condition~\cite{Louis:2012nb} 
\begin{align}
   0\leq   N_{\rm flux}\leq 138.
   \label{eq:tadpole}
\end{align}

By solving SUSY conditions $D_IW=0$ with $I=S,u^1,u^2$, we find that CP-conserving solutions are realized to 
satisfy
\begin{itemize}
    \item $\gamma=0$
{\small
\begin{align}
{\rm Re}u^1&={\rm Re}u^2={\rm Re}S=0,
\nonumber\\
\frac{h_0}{f^0} &= -\frac{{\rm Im}u^1 (3 ({\rm Im}u^1)^2 + 3 {\rm Im}u^1 {\rm Im}u^2 + ({\rm Im}u^2)^2)}{2{\rm Im}S},
\nonumber\\
    \frac{h^1}{{\rm Im}u^1} 
    &= -\frac{ 3 f_2 ({\rm Im}u^1 + {\rm Im}u^2) (3 {\rm Im}u^1 + {\rm Im}u^2) - f_1 (3 ({\rm Im}u^1)^2 + 6 {\rm Im}u^1 {\rm Im}u^2 + 2 ({\rm Im}u^2)^2)} {{\rm Im}u^2 (3 {\rm Im}u^1 + {\rm Im}u^2) (3 ({\rm Im}u^1)^2 + 3 {\rm Im}u^1 {\rm Im}u^2 + ({\rm Im}u^2)^2) {\rm Im}S},
    \nonumber\\
    h^2 &= -\frac{3 f_1 ({\rm Im}u^1)^2 ({\rm Im}u^1 + {\rm Im}u^2) - 
 f_2 (9 ({\rm Im}u^1)^3 + 9 ({\rm Im}u^1)^2 {\rm Im}u^2 + 3 {\rm Im}u^1 ({\rm Im}u^2)^2 + ({\rm Im}u^2)^3)}{{\rm Im}u^1 {\rm Im}u^2 (3 ({\rm Im}u^1)^2 + 3 {\rm Im}u^1 {\rm Im}u^2 + ({\rm Im}u^2)^2) {\rm Im}S},
\end{align}
}
    \item $\gamma=\pi$
{\small 
\begin{align}
{\rm Re}u^1&={\rm Re}u^2={\rm Re}S=0,
\nonumber\\
\frac{f_0}{h^0} &= \frac{{\rm Im}u^1{\rm Im}S \left(3 ({\rm Im}u^1)^2 + 3 {\rm Im}u^1 {\rm Im}u^2 + ({\rm Im}u^2)^2\right)}{2},
\nonumber\\
    \frac{f^1}{{\rm Im}u^1} 
    &= -\frac{ 3 f^2{\rm Im}u^1 ({\rm Im}u^1 + {\rm Im}u^2) (3 {\rm Im}u^1 + {\rm Im}u^2) +2h_1{\rm Im}S (3 ({\rm Im}u^1)^2 + 3 {\rm Im}u^1 {\rm Im}u^2 + ({\rm Im}u^2)^2)} {(3 {\rm Im}u^1 + {\rm Im}u^2) (9 ({\rm Im}u^1)^3 + 9 ({\rm Im}u^1)^2 {\rm Im}u^2 + 3 {\rm Im}u^1 ({\rm Im}u^2)^2+ ({\rm Im}u^2)^3)},
    \nonumber\\
    \frac{h_2}{{\rm Im}u^1}  
    &= \frac{-f^2 {\rm Im}u^2 (3({\rm Im}u^1)^2 +{\rm Im}u^1{\rm Im}u^2+({\rm Im}u^2)^2) +3h_1{\rm Im}u^1 ({\rm Im}u^1 + {\rm Im}u^2){\rm Im}S}{ (9 ({\rm Im}u^1)^3 + 9 ({\rm Im}u^1)^2 {\rm Im}u^2 + 3 {\rm Im}u^1 ({\rm Im}u^2)^2 + ({\rm Im}u^2)^3) {\rm Im}S}.
\end{align}
}
\end{itemize}
On the other hand, we numerically searched\footnote{We used the same numerical method in Sec.~\ref{subsec:4_3}, but the result could be highly dependent on initial values. 
Hence, this non-existence of CP-breaking vacua for the even polynomial case could not be a general statement but may capture some tendency.} for the CP-breaking vacua 
under the set of randomly generated fluxes satisfying the 
tadpole cancellation condition (\ref{eq:tadpole}). 
For the even polynomial case, one cannot find the CP-breaking vacua 
for $4\times 10^7$ dataset of fluxes within $-30 \leq \{f_0, f^{1,2},h^0,h_{1,2}\}\leq 30$, whereas there exist CP-breaking 
vacua in the superpotential with odd degrees. 
Indeed, the numerical search under the randomly generated $2 \times 10^7$ dataset of fluxes
within $-30 \leq \{f^0, f_{1,2},h_0,h^{1,2}\}\leq 30$ leads to 559 stable CP-breaking vacua in the large complex structure regime ${\rm Im}u^1 > 1$. 
For the benchmark dataset of fluxes:
\begin{align}
    (f^0,f_{1}, f_{2},h_0,h^{1}, h^{2})=(-1, 25, -23, 22, 1, -3),
\end{align}
leading to $N_{\rm flux}=116$, 
the vacuum expectation values of moduli fields are evaluated as 
\begin{align}
    &\langle{\rm Re}u^1\rangle \simeq 1.86,\quad
    \langle{\rm Re}u^2\rangle \simeq -2.70, \quad
    \langle{\rm Re}S\rangle \simeq 4.32, 
    \nonumber\\
    &\langle{\rm Im}u^1\rangle \simeq 3.01,\quad
    \langle{\rm Im}u^2\rangle \simeq 4.15, \quad
    \langle{\rm Im}S\rangle \simeq 4.81, 
\end{align}
showing that CP is spontaneously broken due to the nonvanishing values of 
axionic fields.\footnote{We deal with the cubic-type prepotential realized in the large complex structure and weak coupling regime to check whether the CP symmetry is spontaneously broken. 
Even if we include the sub-leading terms to the prepotential as calculated in Ref. \cite{Candelas:1993dm}, the moduli fields are still around these CP-breaking vacua, but the CP symmetry is explicitly broken by the sub-leading effects.}
Note that masses squared of the moduli fields are positive as shown in the descendent order,
\begin{align}
{\cal V}^{-2}(9.72, 5.90, 4.73, 2.38, 5.29\times 10^{-1}, 6.60\times 10^{-2}).
\end{align}

Hence, CP is spontaneously broken in this class of flux compactification. 
The realization of spontaneous CP violation depends on the structure of 
the prepotential and the functional form of the superpotential. 
The reason why the CP-breaking vacua are absent in the choice of the 
superpotential with even degrees is unclear owing to the fact 
that the SUSY conditions are non-linear functions of the axionic fields as in 
Eqs.~(\ref{eq:SUSYeqsodd}) and (\ref{eq:SUSYeqseven}). 
A profound understanding of the origin of CP-breaking vacua will be reached 
by studying other CY flux compactifications, which will be investigated in future work. 
So far, we have focused on the stabilization of the complex structure moduli 
and the axio-dilaton. 
The stabilization of K\"ahler moduli would be realized by non-perturbative 
effects such as D-brane instanton effects. Combining the stabilization 
of K\"ahler moduli with CP-breaking flux compactifications will  also be important future work.

\section{Conclusions}
\label{sec:con}

We have revealed the geometrical origin of CP embedded into the 10D proper Lorentz 
transformation with an emphasis on the complex structure of compact 
6D spaces, in particular the CY threefolds. 
We find that the anti-holomorphic involution of the complex structure is regarded 
as the anti-holomorphic involution of period integrals on CY threefolds 
with the large complex structure. 
Consequently, the anti-holomorphic involution of the period integrals corresponds to 
the outer automorphism of $Sp(2h^{2,1}+2,\mathbb{Z})$ symplectic modular group, 
rather than the element of $Sp(2h^{2,1}+2,\mathbb{Z})$ in the complex structure moduli space. 
The moduli group is then enlarged into $Sp(2h^{2,1}+2,\mathbb{Z})\rtimes {\cal CP}$. 
That is a natural extension of the known toroidal cases, where the CP symmetry is regarded 
as the outer automorphism of the $SL(2,\mathbb{Z})$ modular group~\cite{Nilles:2018wex,Novichkov:2019sqv}.

The CP violation is strongly correlated with the dynamics of the moduli fields, 
whose vacuum expectation values determine the size of CP violation. 
It is an important issue to check whether the CP symmetry is spontaneously broken in 
the string landscape. Indeed, it was stated in Ref. \cite{Kobayashi:2020uaj} that the spontaneous CP violation is difficult to realize on toroidal backgrounds.
To resolve this issue, we examined CY flux compactifications. 
For concreteness, we deal with Type IIB flux compactifications, where 
we turn on three-form fluxes on CY three-cycles. 
Imposing CP invariance on the moduli effective action requires the restricted choices of RR and NSNS flux quanta 
in the large complex structure regime of CY threefolds. 
It results in the flux-induced superpotential consisting of either odd or even 
polynomials with respect to the moduli fields. 
The CP-invariant superpotential possesses the discrete $\mathbb{Z}_2$ symmetry 
or $\mathbb{Z}_4$ R-symmetry from the field theoretical point of view. 
We analyzed the vacuum structure of CP-invariant effective action. 
It turned out that CP-conserving vacua appear in generic CY flux compactifications 
in the large complex structure regime. 
To check the existence of CP-breaking vacua, we work with some concrete CY threefolds. 
It indicates that the prepotential having the toroidal structure does not  lead to the spontaneous CP violation. 
For a particular choice of fluxes, CP is embedded into the duality symmetry of the axio-dilaton and/or 
the modular symmetry of the complex structure moduli. 
When the structure of the prepotential differs from the toroidal one, 
the spontaneous CP violation 
can be achieved on the CY, illustrated on $\mathbb{CP}_{11169}[18]$. 

In this paper, we examined the geometrical origin of CP and its violation in the complex structure moduli space of CY threefolds, but 
phenomenologically it is required to consider the CP violation in the matter sector. 
From our numerical search within $2\times 10^7$ data set of fluxes, 
it turned out that the CP symmetry was spontaneously broken at only $559$ number of vacua, at which the vacuum expectation values of axions as well as saxions are determined. 
Remarkably, these moduli fields appear in gauge kinetic functions and Yukawa couplings of certain D-branes in Type IIB string theory, thereby one can predict the size of strong CP phase as well as the Cabbibo-Kobayashi-Maskawa phase from such a small corner of string landscape. 
Furthermore,  it is interesting to examine more examples to reveal the nature of spontaneous CP 
violation in more broad classes of the string theory, which leaves us for future work.
Also, it was stated in Ref. \cite{Baur:2019kwi} that the CP and the flavor symmetries can be unified in the common group. 
We will report the analysis of CP and the flavor 
on curved CY manifolds in a separate paper.

{\bf Note added}

After finishing this work, Ref.~\cite{Ding:2020zxw} appeared, where symplectic modular symmetries were studied 
from the phenomenological viewpoints.

\subsection*{Acknowledgements}

T. K. was supported in part by MEXT KAKENHI Grant Number JP19H04605. H. O. was
supported in part by JSPS KAKENHI Grant Numbers JP19J00664 and JP20K14477.

\end{document}